\documentclass[12pt]{iopart}
\usepackage{iopams} 
\usepackage{graphicx}
\usepackage{amssymb}
\usepackage{color}

\DeclareMathAlphabet{\mathbi}{OT1}{cmr}{bx}{it}
\SetMathAlphabet{\mathbi}{normal}{OT1}{cmr}{bx}{it}

\newcommand{\zb}{\beta}

\newcommand{\zl}{\lambda}

\newcommand{\zt}{\tau}

\newcommand{\be}{\begin{equation}}
\newcommand{\bea}{\begin{eqnarray}}
\newcommand{\ee}{\end{equation}}
\newcommand{\eea}{\end{eqnarray}}

\begin{document}

\title[Optimal protocols and the Jarzynski equality]
{Applicability of optimal protocols and the Jarzynski equality}

\author{Stuart J.\ Davie, Owen G. Jepps}
\address{Queensland Micro- and Nanotechnology Centre, Griffith University, Brisbane, QLD 4111, Australia}
\address{School of Biomolecular and Physical Sciences, Griffith University, Brisbane, QLD 4111, Australia}
\author{Lamberto Rondoni}
\address{Dipartimento di Scienze Matematiche, Politecnico di Torino,
C. Duca degli Abruzzi 24, 10129 Torino. 
INFN, Sezione di Torino, Via P. Giuria 1, 10125 Torino, Italy}
\author{ James C.\ Reid, Debra J.\ Searles}
\address{Australian Institute of Bioengineering and Nanotechnology, Brisbane, QLD 4072, Australia}
\address{School of Chemistry and Molecular Biosciences, Brisbane, QLD, 4072, Australia}

\begin{abstract}

The Jarzynski Equality is a well-known and widely used identity, relating the free energy difference between two states of a system 
to the work done over some arbitrary, nonequilibrium transformation between the two states. Despite being valid for both stochastic 
and deterministic systems, we show that the optimal transformation protocol for the deterministic case seems to differ from that 
predicated from an analysis of the stochastic dynamics. In addition, it is shown that for certain situations, more dissipative 
processes can sometimes lead to better numerical results for the free energy differences.  

\end{abstract}


\maketitle

\section{Introduction}
The fluctuation relations developed in the past twenty years have opened a new perspective on small driven 
and relaxing systems \cite{ECM,earlyES,GC,Crooks99,Crooks00,ESRJSP,transientexper,expt,Ritort06}. In particular, 
the relation known as the Jarzynski Equality (JE) \cite{Jarz,Jarzynski11} has become popular in various fields,
biophysics in particular. It relates the work $W$ done on a collection of systems driven away from a canonical equilibrium 
state characterized by a parameter $\zl(0)=A$, according to a given protocol $\zl(t)$ which ends at $\zl(\tau)=B$, 
to the free energy difference between the initial state and the one to which the systems may eventually relax:
\begin{equation} 
\left\langle e^{-\zb W} \right\rangle = e^{-\zb \left[ F(B) - F(A) \right]} 
\label{JR} 
\end{equation} 
where $\zb = 1 / k_{_B} T$, $\langle ... \rangle$ is the average with respect to the initial ensemble,
$\left[ F(B) - F(A) \right]$ is the free-energy difference between the initial 
equilibrium state with $\zl=A$ and the equilibrium state corresponding to $\zl=B$. 
Sufficient conditions for Eq.(\ref{JR}) to hold are that the dynamics are reversible, that the system exhibits a 
type of ergodic consistency \cite{Evans03}.  
If these conditions are met, Eq.(\ref{JR}) will hold arbitrarily far from equilibrium (arbitrarily large $d\zl/{d t}$). 

The process always begins in the same macroscopic state, the equilibrium state with $\zl=A$, but in 
different microscopic states. This is why the measured work varies, unless the protocol is quasi-static:
if the protocol is fast enough, the different initial conditions will result in different interactions with
the driving environment, hence in different amounts of work done. At time $\zt$, when the protocol stops, the system will
not in general be in equilibrium. Obviously, these quantities of work are path functions since they depend on the initial 
microstate --- but they are nevertheless measurable quantities \cite{BPRV}.

To build the statistics of the work done, the same protocol must be repeated very many times. Indeed, the left 
hand side of Eq.(\ref{JR}) can be hard to compute precisely in some cases because of its exponential form and the fact that large contributions to that average can sometimes be given by rare, large negative values of $W$ \cite{Jarzynski06}.
Therefore, the question of protocols that optimally achieve the best estimate for the free-energy difference
has been investigated intensely. In order to obtain mathematical results,
most investigations have focussed on stochastic processes, for which numerous techniques are available.
In particular, the works by Schmeidl and Seifert \cite{SchSei} and by Aurell, Mej\'ia-Monasterio and 
Muratore-Ginanneschi \cite{AMM} consider Langevin models such as:
\be
\dot{\xi}_t = -\frac{1}{\zt} \partial_{\xi_t} V(\xi_t,t) + \sqrt{\frac{2}{\zt \zb}} \dot{\omega}_t 
\ee
and come to the conclusion that the optimal protocols at fixed finite $\zt$ are not continuous but are characterized 
by sudden jumps at the beginning and at the end of the process.\footnote{Similarly discontinuous optimal  protocols 
are common, see e.g.\ Ref.\cite{Esposito2010}.} This is consistent with the overdamped nature 
of the evolution equation, but these jumps will disappear for protocols where there is some form of resistance to 
rapid change in the system \cite{Aurell}. The question of the range of applicability of the optimal protocols obtained 
from Langevin-type processes is whether these situations are realizable in practice and, if so, how common they are.

To investigate this question in the case of particle systems, it seems appropriate to consider deterministic 
models, such as those of molecular dynamics, rather than other stochastic processes, which are based on the
same assumptions of the Langevin models. In particular, molecular dynamics models can interrogate the wide 
separation between microscopic scales concerning the constituents of heat baths and systems of interest 
and the mesoscopic scales, which is assumed by the stochastic description. In this sense, the deterministic 
description complements the stochastic one \cite{RMM}.

Here, we consider the problem of fast expansion or compression of a gas by means of an adiabatic piston. 
The gas is initially in equilibrium at a given temperature $T$. It is then isolated from the outer environment, and
the piston starts to move according to a specified protocol. Multiple molecular dynamics simulations were completed 
with various piston velocities. Each simulation modelled a 16-particle system in a 3-dimensional box, with one 
moving wall. The equations of motion for the particles in the system as they move between walls are:

\begin{equation}
\dot{\bf{q}}_{i}=\frac{\bf{p}_{i}}{m}\\
\dot{\bf{p}}_{i}={\bf{F}}_{i}-S_i\alpha \bf{p}_{i}
\label{EOM}
\end{equation}
where $\bf{q}_{i} $ and $\bf{p}_{i} $ are the coordinates and momenta of the $i$th particle, $\bf{F}_{i}$ is the interparticle force on a particle, obtained from a Weeks-Chandler-Anderson short-ranged repulsive pair potential \cite{WCA}, and $S_i$ is a switch to determine whether or not a thermostat is applied.  The thermostat multiplier $ \alpha $ is a Nos\'e-Hoover thermostat \cite{EM}

\begin{equation}
\dot{\alpha}=Q_s \left( \frac{2K}{3Nk_B} -T \right)
\label{NHthermostat}
\end{equation}
where $Q_s$ is a factor that controls the oscillations in the kinetic energy, $N$ is the number of particles in the system, $k_B$ is Boltzmann's constant, $K$ is the instantaneous 
kinetic energy of the system and $T$ is the Nos\'e-Hoover target temperature. In this work, in order to generate initial phase points from a Nos\'e-Hoover canonical distribution \cite{EM}, $S_i=1$ for all particles during the equilibration period.  In order to carry out an adiabatic expansion/contraction, $S_i=0$ for all particles when the piston is moving. Particle-wall interactions reversed the momentum perpendicular to the walls, while 
particle-piston interactions changed the momentum such that $v_{i,z}' = 2 v_{p} - v_{i,z}$, where $z$ 
is the axis parallel to the piston velocity $v_p$, and $v_{i,z}$ and $v_{i,z}'$ are the 
projections on the $z$-direction of the velocity of particle $i$ before and after the collision, respectively.  All results are given in Lennard-Jones reduced units.
The driving protocol consists of either an adiabatic expansion 
from density $\rho(0)=0.1$ to $\rho(\tau)=0.05$, or the opposite (a compression) \cite{DRS}. The protocol involves varying the volume 
of the box by varying the position of the piston in time.  If this process took place quasi-statically, and the object was macroscopic, the
final equilibrium state would be that of a system in an adiabatic container of given volume and energy. 

Because no heat is exchanged with the environment, the work carried out over the period $\tau$ for each trajectory for this system, 
appearing in Eq.(\ref{JR}), is:

\begin{equation}
W=H({\bf q}(t),{\bf p}(t);\lambda=B)-H({\bf q}(0),{\bf p}(0);\lambda=A)
\label{work}
\end{equation}
where $H$ is the internal energy of the system of particles. This is different from the situation described by the Langevin equation 
where the bath is always exchanging energy with the system of interest.  
It is a similar situation to that considered by Bena et al. \cite{Bena}, 
although they consider a deterministic hard sphere system so the change 
in the internal energy is solely due to collisions with the piston.

\begin{figure}
\begin{center}
{\includegraphics[width=7.5cm]{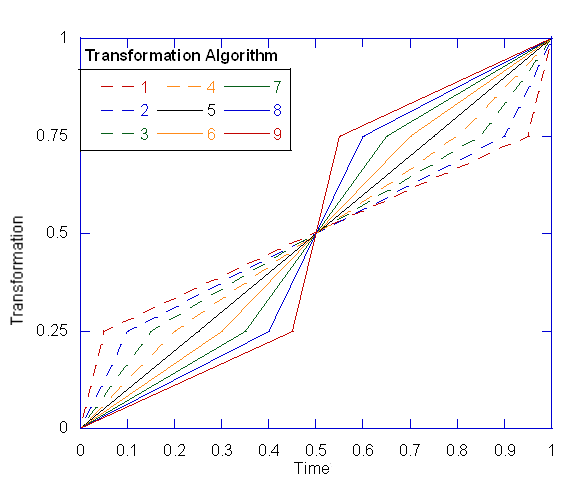}}
{\includegraphics[width=7.5cm]{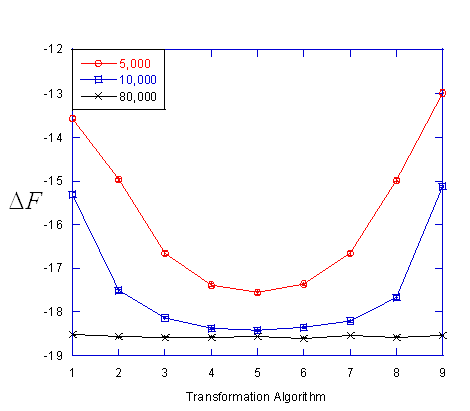}}

\scriptsize{\hspace{0.75cm} (a) \hspace{7.25cm} (b)}
\caption{(a) Graphical description of the transformation algorithms.  In protocols 1-4, the piston velocity is faster at the beginning and end; 6-9 the piston is faster in the middle. Protocol 5 corresponds to a constant piston velocity. (b) Maximum Likelihood Estimator value of $\Delta F$ for the 9 different protocols. The curves represent results for $1.5\times10^5$ simulation runs with $\tau=5000, 10000$, or $80000$ reduced-time units. The standard errors obtained were determined from averaging over 10 groups of $1.5\times10^4$ forward and reverse runs and are smaller than the symbols.}
\end{center}
\label{fig1}
\end{figure}

The Jarzynski Equality \cite{Jarz,6} and the Maximum Likelihood Estimator (MLE) \cite{3,4} here have been used to calculate the 
free-energy differences due to expansion and contraction of a system similar to that considered in Ref.\cite{DRS}.  In the MLE, 
the free energy $\Delta F$ is determined iteratively through solution of the equation
$$
\sum_{i=1}^{N_f} \left( 1 + \frac{N_f}{N_b}\exp(\beta(W_{i,f}-\Delta F)) \right)^{-1} = \sum_{i=1}^{N_b} \left( 1 + \frac{N_b}{N_f}\exp(\beta(W_{i,b}+\Delta F)) \right)^{-1},
$$
where the summations are over the $N_f$ forward and $N_b$ backward trajectories that transform the system 
between the two states at temperature $T= (k_B \beta)^{-1}$ with work $W_{i,f}$ or $W_{i,b}$. Nine different protocols $\zl(t)$,
$t\in [0,\tau]$,  were considered, as illustrated in Figure \ref{fig1}a.  The value of $\lambda(t)=L_z(t)/(L_z(\tau)-L_z(0))$ varies between 0 and 1, where $L_z(t)$ is the position of the piston.  
Changing from 9 to 1, these protocols provide a better and better approximation to the discontinuous jumps described 
by Refs.\cite{SchSei,AMM}.
Each run was repeated $1.5\times10^5$ times for a given simulation length, and the results for the computed
free-energy difference $\Delta F$ have been averaged over 10 groups of $1.5\times10^4$ samples, and reported in Figure \ref{fig1}b,
where the standard error represents their reproducibility, not the error in convergence. As expected for 
quasi-static (sufficiently slow) transformations, $\Delta F$ does not depend on the protocol.
Consequently, the black curve gives a good numerical estimate of $\langle \exp -\zb W \rangle$ in all cases.
However, if the protocol is fast the optimal
protocol appears to be number 5, i.e.\ the smoothest of all, at variance with the results of Refs.\cite{SchSei,AMM}.  
In this case, our model provides one simple example
in which the predictions for the most efficient protocol obtained using the stochastic approach fail.

Our molecular dynamics simulations also provide insight into the nature of minimising the work along the
transformation and the corresponding effects on the calculation of free energies. It 
is generally assumed that an algorithm that minimises the work for a given transformation will also 
generate the most accurate free-energy change calculations, because of the relation
\be
W_{\rm diss} = W - \Delta F
\ee
where $W$ is the work defined above, $\Delta F$ is the free-energy
variation between initial and final states connected by the transformation, and $W_{\rm diss}$ is the
dissipated energy.  This has been demonstrated in some cases, see \cite{SchSei,Vaikuntanathan08,Then08} for example.

\begin{figure}
\begin{center}
\includegraphics[width=7.5cm]{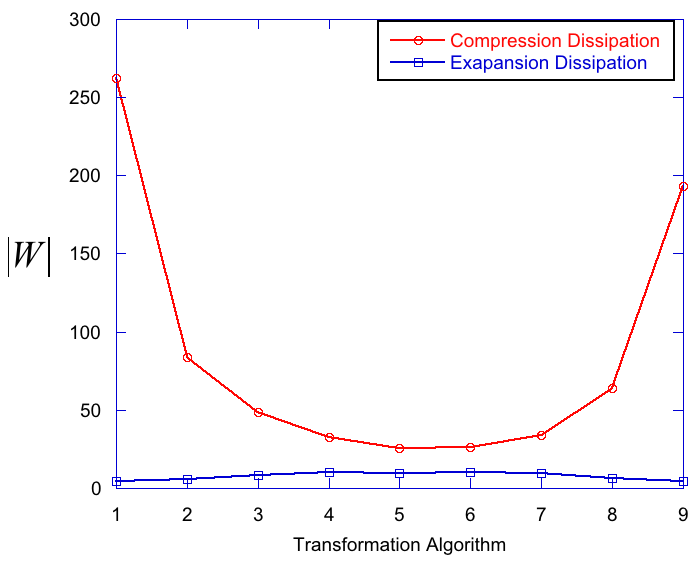}
\includegraphics[width=7.5cm]{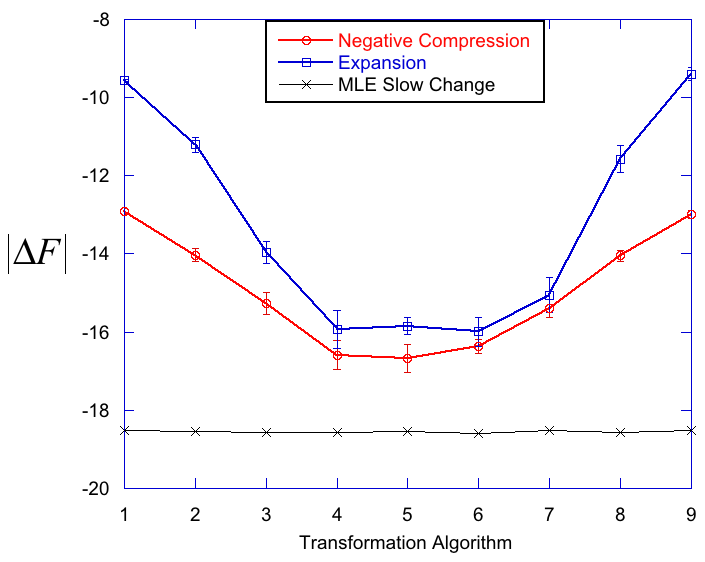}

\scriptsize{ \hspace{1cm} (a) \hspace{7.25cm} (b)}
\end{center}
\caption{Comparison of (a) dissipation and (b) free-energy differences for expansion and compression protocols. 
Despite the dissipation associated with the expansion process being less than that associated with the 
compression process, the compression simulations return more accurate free-energy differences.}
\label{freeenergies}
\end{figure}

Figure \ref{freeenergies}a displays the average dissipation obtained from the 5,000-timestep, variable piston-speed simulations for 
both expansion and compression simulations. Figure \ref{freeenergies}b displays the free-energy differences obtained from the Jarzynski Equality for these same transformations, and compares them to the 
slow-change result. 
The negative of the compression free-energy change is shown for ease of comparison. 
Despite the significantly lower dissipation obtained from the expansion simulations, the compression simulations 
produce more accurate free-energy calculations. When only considering a one-directional transformation, 
the algorithm which minimises dissipation also produces the most accurate free-energy differences. Out 
of the expansion protocols, the one with the greatest dissipation provides the best estimate of the 
$\Delta F$.
Thus, because the free-energy 
difference between two states can be calculated in either direction, this demonstrates that a more 
dissipative transformation path can produce more accurate free energies than a less dissipative one.

In our work we also obtained the distributions of $W$ for the rapid expansion and compression, which  
vary quantitatively and qualitatively with the protocol.  

\begin{figure}
\begin{center}
\includegraphics[width=7.5cm]{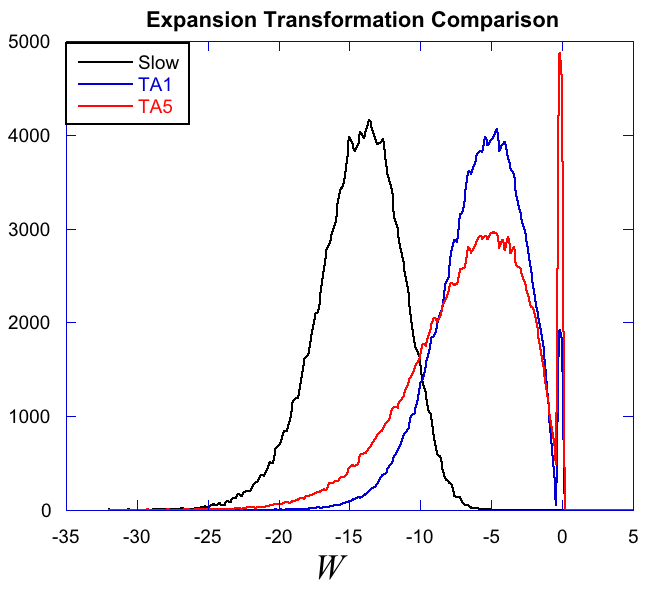}
\includegraphics[width=7.5cm]{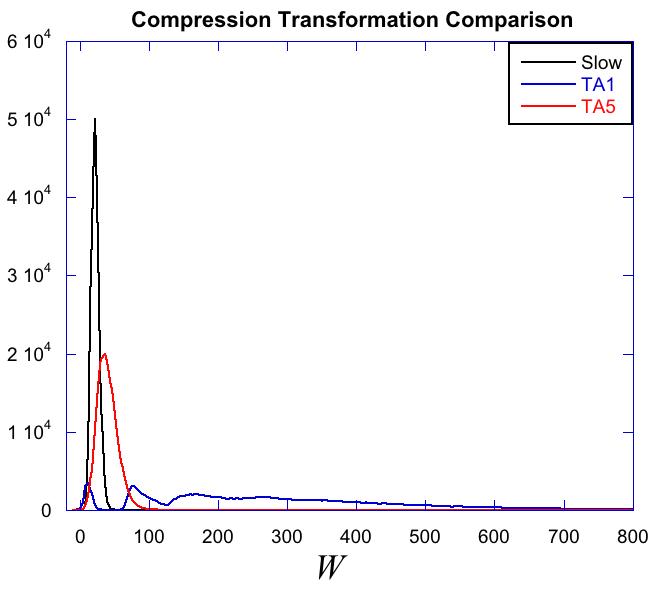}

\scriptsize{\hspace{0.75cm} (a) \hspace{7.25cm} (b)}
\end{center}
\caption{Distributions of the work for a) expansion and b) compression processes for the slow change, and protocols 1 and 5.}
\label{distributions}
\end{figure}

As shown in Figure \ref{distributions}a, the distributions for the rapid expansion possess two separate 
peaks, with the narrower one centred close to zero, corresponding to transformations without any 
particle-piston interactions. 
This causes a significant sampling problem that is different to the stochastic case, since for a very fast expansion almost no particles will be moving fast enough in the deterministic simulation
to interact with the piston in each trajectory, and the work obtained will be close to zero for almost all trajectories.  This results in a substantial 
 number of very small work values observed in the work distribution, giving a value for the average work that is closer to zero than is obtained with a slow change.  Furthermore, if insufficient trajectories are used for the numerical calculations, the sampling of larger negative values of the work will be poor, resulting in free-energy differences (calculated via the finite-sample version of Eq.(\ref{JR})) that differ from the correct $\Delta F$.  We emphasize that this is a sampling problem, not a theoretical problem with Eq.(\ref{JR}) (see \cite{Lua05} for a discussion).   
 
Figure \ref{distributions}b shows the work distributions from rapid compression. Protocol 1 leads to a very irregular 
distribution, while protocol 5 gives quite a regularly shaped distribution, and the slow protocol yields 
a distribution that is sharp initially, and then smooth. However, the results obtained using protocol 1 for compression are noticeably more accurate than those obtained from the expansion work data using the reverse protocol. This is because phase space has been sampled more completely in areas of the distribution that contribute significantly to the ensemble average in Eq.(\ref{JR}). 
The peaks in the work distribution for a system undergoing rapid compression are due to particle-piston 
interactions, as each particle-piston interaction will amount to a significant contribution to the total work. 
In fact, in the limit of an infinitely fast piston, the average number of particle-piston interactions will
become a function solely of the ratio of the pre- and post-compression system lengths.  High-speed compressions
may significantly increase the mean work (due to the high-energy piston-particle interactions), however the exponentially 
weighted mean is much less sensitive to these trajectories.  Therefore, although the compression results in significantly 
more dissipation, the convergence of the free-energy determined by the Jarzynski equality is better. 

Our work, which was performed with a system of only 16 particles, highlights the sampling issues that can result when 
thermodynamic properties are determined as ensemble averages of mechanistic experiments.  In cases of few degrees of 
freedom, elements of the mechanics can dominate the  behaviour expected with many particles.  For example, the initial 
microscopic conditions are taken from the canonical distribution, but the speed of the protocol affects the final set 
of states: as the expansion protocol becomes faster, the final state is reached with less and less work on average 
being done on the piston.  As noted, what we do is different to the case described by the Langevin equation because 
no bath is acting.  However, if we introduce a Nos\'e-Hoover thermostat, our results, and particular the optimal 
protocol, do not seem to change qualitatively.

We can compare the work distributions with those obtained by Bena et al. \cite{Bena} for deterministic hard 
sphere systems. They obtained analytical results for a Jepsen gas and compared them with numerical simulations 
of a dilute hard sphere gas.  The rate of expansion/contraction was kept constant in all cases, and they compared 
the work distributions for different rates. Like in the current work, asymmetry of the work distributions for 
the compression and expansion protocols is observed and the distributions become less Gaussian as the rate is 
increased.\footnote {Note that $W$ is the work done 
by the system in \cite{Bena}, whereas it is work done on the system in our case, so the signs are reversed} 
By comparing the analytical results for the infinite systems with the numerical results they also observed effects 
due to the use of a finite number of particles in the simulations, which would be even more pronounced in our 
systems which are two orders of magnitude smaller than those of \cite{Bena}. In contrast with our results which 
are for particles interaction via soft repulsive potentials, the particles are non-interacting and this will also 
have some effects on the result e.g.\ Ref.\cite{Pulvirenti,CohR,FPPRV}. However, the implications of this asymmetry of 
the distributions on the Jarzynski equality are discussed and are consistent with our observations.  
Therefore, should a study of the optimal protocol for this system be undertaken like in our current work, we 
would expect similar results which are unlike the results predicted for the stochastic systems.  

We have performed molecular dynamics simulations to calculate the free energy associated with adiabatic expansions 
and compressions of a low density gas, complementing existing results already obtained using stochastic simulation 
techniques. We observe an optimal protocol for obtaining free-energy estimates using the JE that differs  substantially from the protocol observed for stochastic models, even those allowing accelerations.
We also note a remarkable and unusual feature of our results, that the better free-energy estimates can be associated 
with more highly dissipative processes.  
Finally, the distinction between our scenario and that of the stochastic approach is reinforced by 
papers such as Ref.\cite{Esposito2010} that show beyond the Langevin framework, the impact of the 
separation of scales assumed in the stochastic approach is not necessarily verified in small deterministic systems. \\

\section*{Acknowledgments}
The authors would like to thank the Australian Research Council for support of this project through a Discovery
Project.  Computational resources used in this work were provided by Griffith University, University of Queensland, 
an Australian Research Council LIEF grant  and the Queensland Cyber Infrastructure Foundation. 
LR acknowledges support from the European Research Council under FP7/2007-2013 Framework Programme, ERC grant 
agreement n.\ 202680. The EC is not liable for any use that can be made on the information contained herein.

\end{document}